# Coverage and Connectivity Aware Neural Network Based Energy Efficient Routing in Wireless Sensor Networks

[1]Neeraj Kumar, [2] Manoj Kumar, [3]R.B. Patel

[1,2]School of computer Science and Engineering, SMVD University, Katra (J&K), India

*nehra04@yahoo.co.in, vermamk@gmail.com*

[3]Department of Computer Science and Engineering, MITS University, Sikar, Rajassthan
*patel_r_b@yahoo.com*

### *Abstract*

*There are many challenges when designing and deploying wireless sensor networks (WSNs). One of the key challenges is how to make full use of the limited energy to prolong the lifetime of the network, because energy is a valuable resource in WSNs. The status of energy consumption should be continuously monitored after network deployment. In this paper, we propose coverage and connectivity aware neural network based energy efficient routing in WSN with the objective of maximizing the network lifetime. In the proposed scheme, the problem is formulated as linear programming (LP) with coverage and connectivity aware constraints. Cluster head selection is proposed using adaptive learning in neural networks followed by coverage and connectivity aware routing with data transmission. The proposed scheme is compared with existing schemes with respect to the parameters such as number of alive nodes, packet delivery fraction, and node residual energy. The simulation results show that the proposed scheme can be used in wide area of applications in WSNs.*

## 1. INTRODUCTION

Wireless Sensor Networks (WSNs) is a class of wireless ad hoc networks in which sensor nodes collect, process, and communicate data acquired from the physical environment to an external Base-Station (BS). But these networks have several challenges such as sensor nodes in WSNs are normally battery-powered, and hence energy has to be carefully used in order to avoid early termination of sensors' lifetimes [1]. As such, the concept of continuous monitoring of network resources becomes a very important topic in WSNs. This same concept has been already investigated in many other environments, e.g., power plants [2], and in many distributed systems [3]. Many recent experimental studies have shown that, especially in the field of sensor networks where low power radio transmission is employed, wireless communication is far from being perfect [4,5,6].

WSNS can be used to monitor the interested region using multi-hop communication. Coverage is a primary metric to evaluate the monitoring capacity. Connectivity also should be guaranteed so that BS can receive all sensed data for future processing. The nodes in WSNs have processing and communication capacities, which can collect surrounding information and then transmit report data to BS. BS analyzes the data received and decides whether there is an unusual or exceptional event occurrence in the deployed region. However due to the limited battery power the sensor nodes could not execute complex instructions or algorithms. For many applications, the desired lifetime of a sensor network is of order of a few years. It may be infeasible or undesirable to recharge batteries in sensor nodes once a WSN is deployed. Hence, energy efficiency is a paramount design consideration for all WSNs.

Many efforts have been made for energy efficiency of WSNs, most of which proposed new network protocols to preserve energy [1,7]. Many researchers focused on next hop selection strategies, which made one-hop neighboring nodes or multiple link for transmitting data [8–10]. Some researchers focused on scheduling the nodes by switching the mode of nodes between sleep and work mode to save energy [11–13].

In this paper, we propose coverage and connectivity aware energy-efficient clustering and routing in WSNs. A neural network based clustering mechanism is proposed for selection of cluster head (CH) among the participating nodes. The problem is formulated as LP with specified constraints having an objective of maximizing the network lifetime. We define an efficient routing metric to be used in taking the selection of next hop in routing.

Rest of the paper is organized as follows. Section 2 describes the related work. Section 3 discusses the network and energy models, routing metric, problem formulation and constraints. Section 4 describes the proposed solution supported with algorithms. Section 5 describes the simulation environment and results obtained. Finally, Section 6 concludes the paper.

## 2. RELATED WORK

Energy balancing and network lifetime of WSN have drawn much attention in recent years. Many previous energy efficient mechanisms adopt different means to balance energy consumption among sensor nodes to prolong network lifetime. Energy Aware Routing (EAR) [9] builds multiple paths from data sources to a sink node. Using a stochastic approach, it selects sub-optimal next hops for each node, but it can only gain energy balance locally. There are number of clustering protocols have been proposed in literature e.g. LEACH [14], PEGASIS [10], HEED [15], EEUC [16], and FLOC [17]. The cluster formation overhead of the clustering protocols includes packet transmission cost of the advertisement, node joining and leaving, and scheduling messages from sensor nodes. All these protocols do not support adaptive multi-level clustering, in which the clustering level cannot be changed until the new configuration is not made. Therefore, the existing protocols are not adaptable to the various node distributions or the various sensing area. If the sensing area is changed by dynamic circumstances of the networks, the fixed-level clustering protocols may operate inefficiently in terms of energy consumption. Bandyopadhyay and Coyle [18] proposed the randomized clustering algorithm to organize sensors into clusters in a wireless sensor network. Computation of the optimal probability of becoming a cluster head was presented. Moscibroda and Wattenhofer [19] defined the maximum cluster-lifetime problem, and they proposed distributed, randomized algorithms that approximate the optimal solution to maximize the lifetime of dominating sets on wireless sensor networks. Pemmaraju and Pirwani [20] considered the k-domatic partition problem, and they proposed three deterministic, distributed algorithms for finding large k-domatic partitions. Tan and Korpeoglu [21] proposed two new algorithms under the name PEDAP, which are near optimal minimum spanning tree based wireless routing scheme. The performance of the PEDAP was compared with LEACH and PEGASIS, and showed a slightly better network lifetime than PEGASIS. Yi et. al [22], presents a Power efficient and adaptive clustering protocol PEACH. The problem of scheduling sensor activities to maximize network lifetime while maintaining both discrete K-target coverage and network connectivity is studied in [23]. In this approach, it is required that each target should be simultaneously observed by at least K sensors. The problem of combining the connectivity with coverage with given network coverage ratio under border effects is studied in [24]. In this scheme, the scenario where the sensor nodes are distributed in a circle-shaped region uniformly are studied. They first derive the network coverage provided by N sensor nodes by the mathematical formulae and then lower bound of the network connectivity probability is also derived in [24]. Moreover various cluster head election techniques with coverage preservation are studied in [25].The authors in [26] propose a novel Adaptive State-based Multi-path Routing Protocol (ASMRP), which constructs Directed Acyclic Graphs (DAGs) from each Mesh Router to Gateways and effectively discovers multiple optimal ath set between any given router and gateway pair. A congestion aware traffic splitting algorithm to balance traffic over these multiple paths is presented which improves the overall performance of the WMNs.

# 3. MODELS, ROUTING METRIC AND PROBLEM FORMULATION

## 3.1 Network Model

We consider a network of homogeneous and energy-constrained sensor nodes that are randomly deployed in a sensor field. Sensor nodes are initially powered by batteries with full capacities. Each sensor collects data which are typically correlated with other sensors in its vicinity, and then the correlated data is sent to the BS via Cluster Head (CH) for evaluation or decision making purposes. We assume periodic sensing with the same period for all sensors and CH is elected as in [14]. Inside each fixed cluster, a node is periodically elected to act as CH through which communication to/from cluster takes place [14].

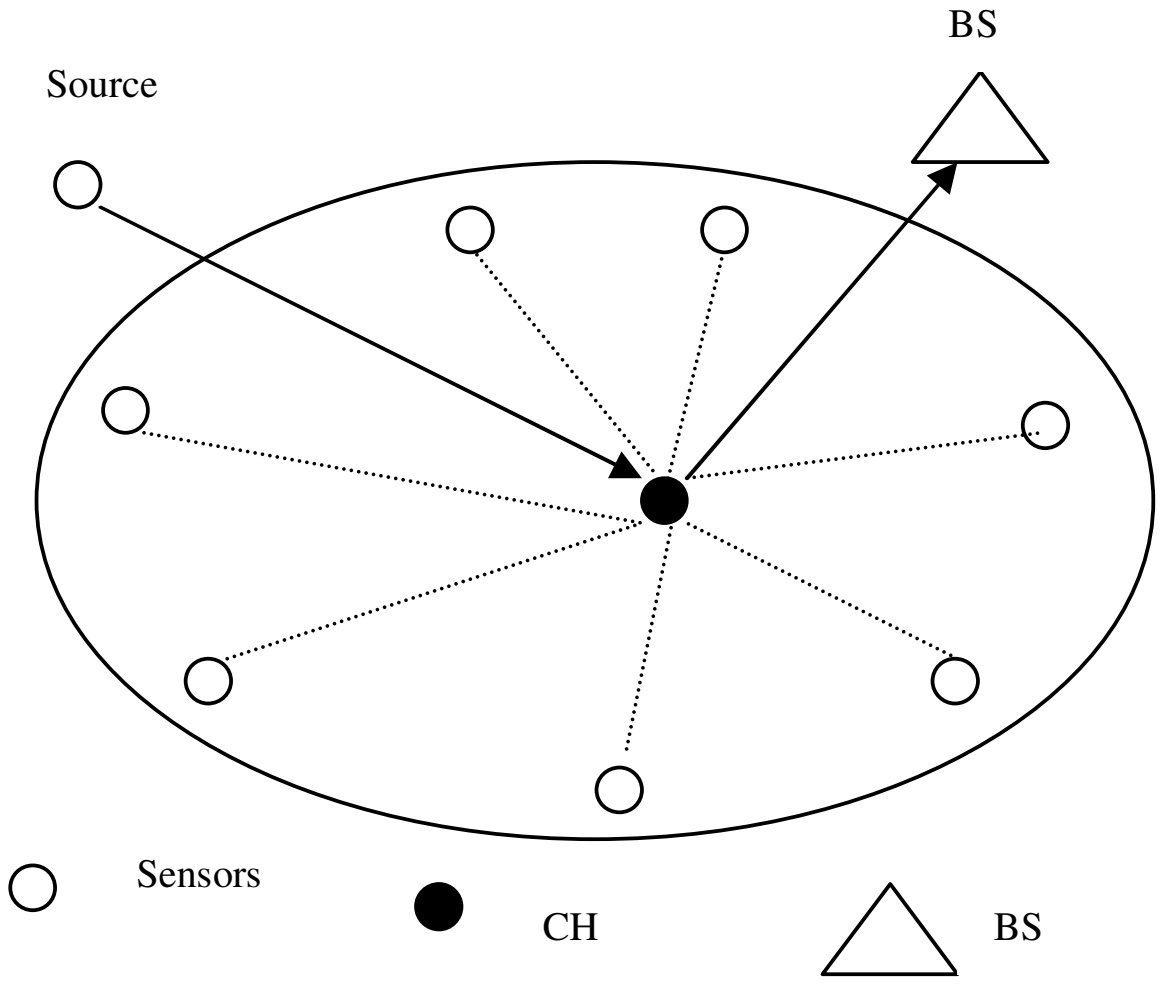

Figure 1. Data Transmission in typical Sensor Networks

## 3.2 Energy Model

To ascertain the amount of energy consumed by a radio transceiver, we apply the following energy model. For each packet transmitted by a sending node to one or more receivers in its neighborhood, the energy is calculated as according to [14]:

$$e = e_t + ne_r + (N - n)e^h{}_r \ldots\ldots\ldots\ldots\ldots\ldots\ldots\ldots(1)$$

where $e_t$ and $e_r$ denote the amount of energy required to send and receive, n is the number of nodes which should receive the packet, and N the total number of neighbors in the transmission range. $e^h{}_r$ quantifies the amount of energy required to decode only the packet header
According to model described in[14], $e_t$ and $e_r$ are defined as

$$e_t(d,k) = (e_{elect} + e_{amp} * d^{\rho})8k$$
$$e_r(k) = e_{elect} * 8k \qquad \text{..................(2)}$$

for a distance $d$ and a $k$ byte message. We have set
$e_{elect} = 70nJ/bit$, $e_{amp} = 120pJ/bit/m^2$, $d = 60m$, $\rho = 4$

### 3.3 Routing Metric

The cost of a link between two nodes $S_i$ and $S_j$ is equal to the energy spent by these nodes to transmit and to receive one data packet, successfully. To establish the coverage and connectivity aware connection between two sensors, a proper routing metric is needed which will guide to form the connection between the sensors. The following routing metric $R_C$ is chosen and calculated as follows:

$$R_C = \left( \frac{E_i^{\,D}}{E_t(S_i,S_j) + E_r(S_i,S_j)} * \frac{1}{R} \right), \text{ ..........(3)}$$

where $E_i^{\,D}$ is energy associated with the delivery ratio of the packet originating from source node $S_i$ and correctly received at destination node, while $E_t(S_i,S_j)$ is the energy used in transmitting from $S_i$ to $S_j$ and $E_r(S_i,S_j)$ is the energy used in receiving the packet, R is the total coverage area.

### 3.4 Problem with the Existing Solutions

The potential problem in the existing protocols is that once the optimal route is determined, it will be used for every transmission. This may not be an ideal solution from network's point of view. Using the optimal path frequently leads to energy depletion of the nodes along that path and, in the worst case, may lead to network partition. To counteract this problem, data forwarding could use multiple paths at different times; thus, any single path does not get energy depleted quickly. Compared with the single path strategy, multipath routing can balance traffic loads among multiple nodes, and can respond to network dynamics where nodes can join and leave the network. Chang and Tassiulas [27] proposed an algorithm to route data through a path whose nodes have the largest residual energy. The path is changed adaptively whenever a better path is discovered. The primary path will be used until its energy falls below the energy of the backup path, at which point the backup path is used. In this way, the nodes in the primary path will not deplete their energy resources due to continual use of the same route, thus achieving longer network life. But they did not consider the coverage and connectivity issue with respect to routing so that every node can participate in CH election mechanism. Moreover, the valuable data to be sent by the nodes would not be missed, if coverage and connectivity is preserved.

### 3.5 Problem Formulation

In this section, we will formulate the coverage and connectivity aware routing metric that binds all the nodes operating in a particular region. Let there are $N$ sensors $\{S_1, S_2, ..., S_N\}$ that are randomly deployed to cover an area $A$. Each sensor has initial energy $E_0$ and covers a disk of radius $R$. Sensor consumes energy $E_c^{\,i}$ to communicate with 1-hop neighbors. A sequence of covers $\{C_j, j=1,2,....,K\}$ consists of sensors such that $A \subset \bigcup_{S_i \in C_j} J(S_i)$ for $j=1,2,....,K$. A cover must ensure that every point in the region is within the sensing range of at least one sensor. For a given set of $N$ sensors, there are $2^N$ possible combinations of coverage sub regions by each of the $N$ sensors. These subregions are disjoint to each others. We denote the $M$ disjoint sub regions by $\lambda_m, m=1,2,...,M, \lambda_i \cap \lambda_j = \phi$ and $\bigcup_{m=1}^{M} \lambda_m = A$. Each sub regions have set of sensors nodes $\lambda_m = \{S_1{}^m, S_2{}^m,..., S_N{}'^m\}$, where $N'$ is the number of sensors covering $\lambda_m$, and $\lambda_m \subset J(S_i{}^m), i=1,2,....,N'$.

**Definition:** A point in an area $A$ is said to be covered if it lies within the sensing range of at least one sensor. A *cover* is a subset of the sensors that provides complete coverage of the region, i.e. $A \subset \bigcup_{S_i \in C_j} J(S_i)$.

We have extended the problem and constraints as presented in [28] by T. Zhao and Q. Zhao. The coverage and connectivity criteria presented by T Zhao is not suitable for dynamic creation of network. So a routing metric is needed which will be updated automatically as the node leave or join the network to have a connectivity. Hence to meet the requirements of the dynamic changes in the network, we have defined the additional routing metric in equation (3) for the dynamic environment and modified the criteria to find the effective regions to be selected.

In fact, we have to find the best sequence of covers to maximize the life of the network and minimize the energy consumption. So the problem reduced to $\max imize \quad K$ and $\min imize \ R_metric$ .(i.e. maximum coverage with minimum energy consumption).

Subject to the following constraints

i) Coverage Constraints: $A \subset \bigcup_{S_i \in C_j} J(S_i)$ for $j=1,2,....,K$ ..........(4)

ii) Energy constraints: $\sum E_c{}^i \le E_0$ , for $i=1,2,...,N$ ..............(5)

iii) Maximum Lifetime Constraints: $E_C{}^i \le P_{\max imum}$ .................(6)

.

iv) Amount of data to be transmitted between two sensors: $D_{ij} \ge 0$ , $1 \le j \le n$ .........(7)

v) $\sum_{1 \le j \le n} D_{ij} - \sum_{1 \le j \le n} D_{ji} = b_i$ ............(8)

Constraint (8) specifies the amount of data transmitted $b_i$ between two nodes $S_i$ and $S_j$ .

Constraint (4) ensures the coverage of complete region of the network under consideration, constraint (5) ensures the energy consumption for communication with 1-hop neighbors should be less than initial energy to prolong the network lifetime. Constraint (6) limits the maximum power consumption of any node in the network. Constraint (7) limits the amount of data to be transmitted between sensors to reduce the energy consumption, Constraints (8) specifies the amount of data transmitted $b_i$ between two nodes $S_i$ and $S_j$

# 4. PROPOSED SOLUTION

We have proposed an energy-aware routing protocol that uses a set of suboptimal paths to increase the lifetime of the network. The whole network is partitioned into different disjoint subregions. From these subregions effective cover set is made depending upon the routing metric defined in equation (3) and defined constraints. Sensors are grouped together depending upon their routing metric and residual energy. Each sensor energy is updated automatically after each round.

The proposed solution to above cited optimization problem consists of following steps:

1) Division of multiple sub regions
2) Selection of Cluster head using neural network
3) Selection of optimized route
4) Multi path Data Transmission

## 4.1 Division of multiple sub regions

The algorithm for partition of network in to different sub regions to preserve the coverage and connectivity is described in Figure 2.

**Step1: Creation of Sub regions**

1. Partition the sensing range $R$ and location into $A$ disjoint regions $\lambda_m, m = 1,2,...,M$

2. Associate each sensor with a set of sub regions that it covers such that $\bigcup \lambda_m = J(S_i)$.

**Step2. Formation of Cover set**

Repeat the following steps until all the sensors are not covered

3. find the efficient sub regions as $B^{eff} = \min \sum_{S_i \in \lambda_m} R_C$.

4. From the covering region $B^{eff}$ choose the sensors having minimum value of $R_C$.

6. Among the selected sensors, choose the sensors with maximum value of residual energy $E_r{}^i$

**Step3. Update of residual energy**

7. For all the sensors in the current cover set update their energy as follows:
$E_r{}^i = E_r{}^i - E_c{}^i$

8. If $(E_r{}^i < E_c{}^i)$, then drop the sensor from current cover set

9. Update the values of the sub regions and their relationships.

Figure2. Creation of Sub regions dynamically

## 4.2 Neural Network based Cluster Head Election

Once the whole region is divided into different regions, the next phase is to choose the CH among the participating nodes to balance energy consumption. Many CHs election mechanism are proposed over the years out of which many proposals favor uniformly distributed clusters with stable average cluster sizes [14-18]. But we propose a new neural network based coverage and connectivity aware clustering algorithm. The set of cluster head nodes can be selected based on the routing cost metric defined in equation 3. The densely populated parts of the network will be overcrowded with CH nodes, while the scarcely covered areas will be left without any CH nodes. In such a situation, it is likely that the high cost sensors from poorly covered areas will have to perform expensive data transmissions to distant CH nodes which will further reduce their lifetime. So we propose neural network based approach. There are three layers in the proposed neural network approach: Input layer, Competition layer and Output Layer.

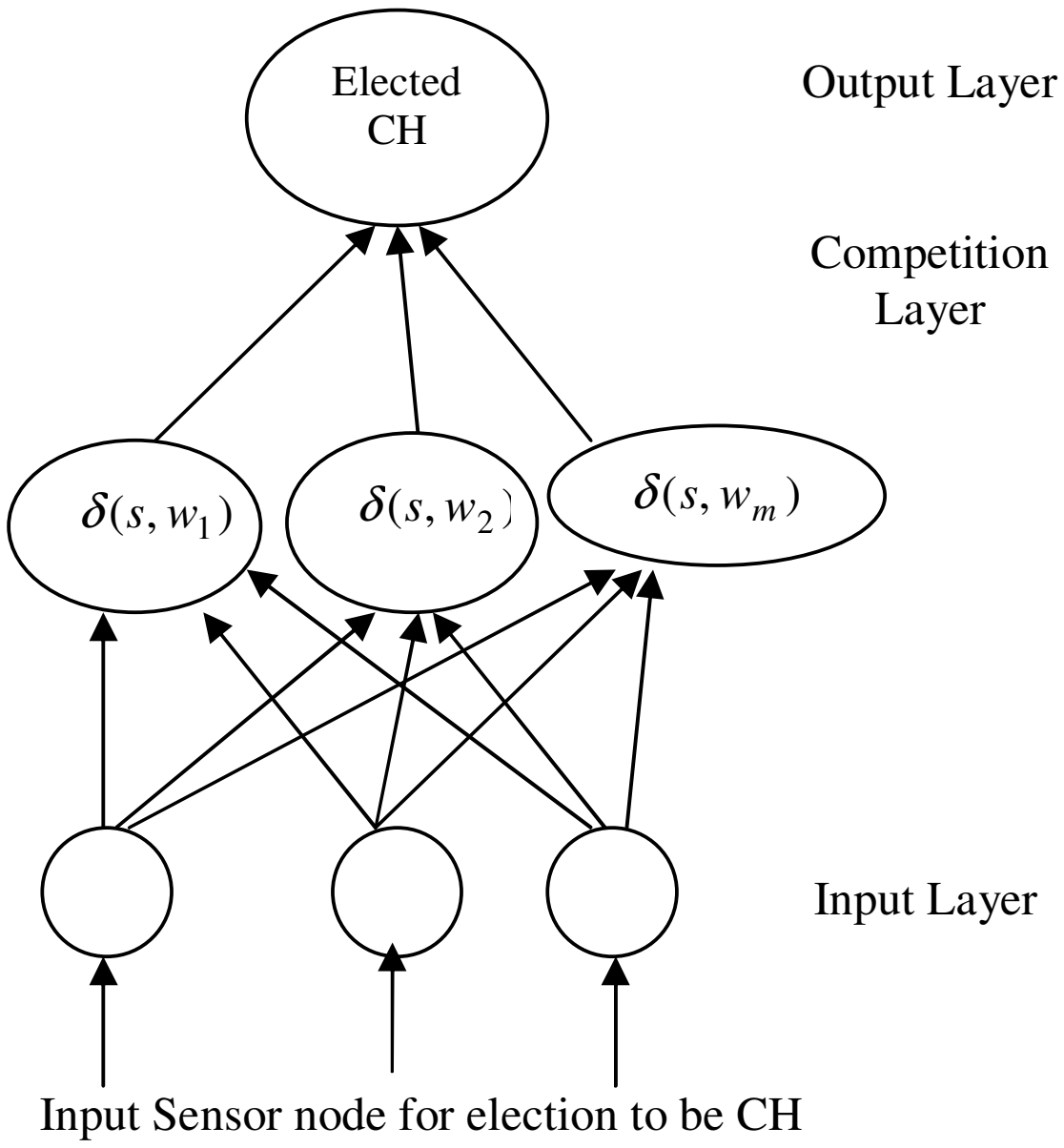

Figure3. Selection of CH

Neural networks have solved a wide range of problems and have good learning capabilities. Their strengths include adaptation, ease of implementation, parallelization, speed, and flexibility. A two - layer feed forward neural network that implements the idea of competitive learning is depicted in Figure 3 above. The nodes in the input layer admit input patterns of sensor nodes competing for CH and are fully connected to the output nodes in the competitive layer. Each output node corresponds to a cluster and is associated with weight $W_j$, $j = 1,2,....,m$, where $m$ is the number of clusters.

The neurons in the competitive layer then compete with each other, and only the one with the smallest $E_i^D$ value becomes activated or fired. Each neuron in the proposed algorithm for CH selection has an adaptive learning. The learning rate $\mu$ determines the adaptation of the vector towards the input pattern and is directly related to the convergence. If $\mu$ equals zero, there is no learning. If $\mu$ is set to one, it will result in fast learning, and the prototype vector is directly pointed to the input pattern. For the other choices of $\mu$, the new position of the vector will be on the line between the old vector value and the input pattern. Generally, the learning rate could take a constant value or vary over time. The algorithm for cluster head election is described in figure 4.

1. Initialize the Vector $S = \{S_1, S_2, ...., S_m\}$ of sensor nodes competing for Cluster head.
   **//Processing at Input Layer**
2. Choose a winner $k$ from sensor nodes as CH whose $E_i^D$ is minimum as follows
   $k = \arg\min\{E_i^D\}$ **// Competition Layer**
3. Also take $Z_i^D$ as the smallest Euclidean distance of sensor node to BS i.e.
   $Z_i^D = k \sum_{i=1,2,...m} |S_i - BS|$, where $k$ is proponality constant
4. Initialize the weight vector as $w_j(old) = Z_i^D$.
5. Update the value of weight vector as follows:
   $w_j(new) = w_j(old) + \mu(S_i - w_j(old))$, where $\mu$ is learning rate of the neurons. $0 \le \mu \le 1$
6. Repeat Steps (2-4) iteratively.
7. Neuron with smallest value of $E_i^D$ is winner. **// Output Layer**

Figure4. Algorithm for Cluster head selection

### 4.3 Selection of Optimized Route

Let us denote by $R$ the maximum number of routes that exist between each source-destination pair, and $l$ is the indication of a route in R. Also, denote by $pow(S_i, l)$ the power consumed by node $S_i$ in transmitting to the next node on route $l$. For the sake of simplicity, we assume that this parameter depends only on the distance between the transmitting and the receiving node. Then, we associate with each route $l$ an energy cost routing metric defined in equation (3) above. The proposed algorithm scans all the routes in R and determines the least expensive route to reach the BS. A source will select the route that has the least energy consumption or the one that maximizes the network lifetime.

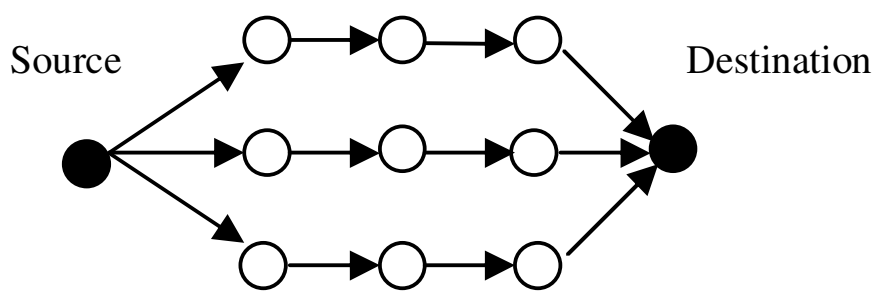

Figure 5. Multipath Routing

To obtain the routes, BS first generates a *Route Discovery* message that is broadcasted throughout the network. Upon receiving the broadcast message, each sensor node introduces a delay proportional to its cost before it forwards the *Route Discovery* message to the nodes in range $R$. In this way a message arrives at each node along the desired minimum cost path. The cumulative cost of the routing path from BS to the node obtained in this phase is taken as energy aware routing *cost as* described in (3). Given $m$ available paths, the overall energy consumption per packet, $E$, can be written as $E = \sum_{i=1,2...m} E_i L$, where $E_i$ is the energy consumption for one bit along path $i$ and $L$ is the packet length in bits. The detailed algorithm for route selection is described in Figure 6.

1. Sort the paths $p_1, p_2, .......p_m$ according to $E_i^{D}$ as $E_i^{D_1} < E_{i+1}^{D_{+11}} < .......E_m^{D_m}$
2. $j = 1$ // initialize the counter for available paths.
3. for each path from source to destination do
4. **repeat**
$E_{C,m} = \sum_{i=1,2...m} m_i E_{\max} L$ // $E_{C,m}$ is use to store the minimal energy consumption per bit with $m$ paths and is assigned maximum value initially.
6. $R_C = 0$ // initialize the value of routing cost
7. **repeat**
8. Solve equation (3) and get the corresponding optimal energy distribution with respect to constraints defined in equations (4),(5),(6),7.
9. Calculate $E_{C,m} = \sum_{i=1,2...m} E_i L$
10. Calculate the value of $R_C$ from equation (3) and $R_C^{updated} = R_C$ // Update value of routing cost
11. **Until** $| R_C^{updated} - R_C | < \delta_1$ (predefined threshold)
12. Update the values of energy for each route discovery process as $E_{C,m}^{Updated} = E_{C,m}$
**Until** $| E_{C,m}^{Updated} - E_{C,m} | < \delta_2$
13. $j = j + 1$ // Update the counter of the paths
14. Compare all paths using $R_C$ metric and select the smallest one.

Figure 6. Algorithm for Route selection

## 4.4 Multipath Data Transmission

This phase is mainly concerned about the multipath data transmission once route selection process is over. The optimal path chosen in section 4.3 is that which consumes less energy and maximum coverage. But the potential problem in the previous defined protocols [14-18] is that once the optimal route is determined, it will be used for every transmission. Using the optimal path frequently leads to energy depletion of the nodes along that path and may lead to network partition. To counteract this problem, we propose multipath data transmission along different paths. These paths are chosen by means of a probability that depends on how low the energy consumption of each path is. Due to the probabilistic choice of routes, it continuously evaluates along different routes and optimal paths are chosen accordingly as defined in section 4.3.

Let $N_i, N_j, N_s, N_D$ are the intermediate, source and destination nodes, while $R_C$ is the routing metric. Initially the value of routing cost $R_C$ is set to zero but it is updated as the data transmission takes place along the optimal path. The algorithm for data transmission is described in Figure 7.

The destination node sends the request in the direction of the source node. It also sets the routing cost field to zero before sending the request as:

1. $R_C(N_D) = 0$ // Set the routing cost zero before sending the request
2. $While(d(N_i, N_S) \geq d(N_j, N_S) \&\&(d(N_i, N_D) \leq d(N_j, N_D))$
3. Send the request only to a neighbor $N_j$
4.For each request sent from node $N_i$ to node $N_j$, calculate the cost of the path as

$$R_C(N_i, N_j) = R_C(N_S, N_i) + R_C(N_i, N_D)$$

5. Discard the paths having high cost and do not include in the forwarding routing table as

$$R_FT_j = \{i \mid R_C(N_j, N_i) \leq \min(R_C_{N_j,N_k})\}, 1 \leq k \leq n$$

6. Assigns a probability to each of the neighbors in the forwarding routing table $R_FT_j$, as follows:

$$P_{j,i} = \frac{1/R_C_{N_j,N_i}}{\sum_j \frac{1}{R_C_{N_j,N_k}}}$$

7. Thus, each node has a number of neighbors through which it can route packets to the destination. Each node then calculates the average cost of reaching the destination using the neighbors in the forwarding table
8. This average cost, $R_C(N_D)$, is set in the cost field of the request packet and forwarded along toward the source node as in the above steps.
9. The source node sends the data packet to any of the neighbors in the routing table, with the probability of the neighbor being chosen equal to the probability in the forwarding table and the same procedure follows for the intermediate node also.

Figure7. Algorithm for data transmission along multipaths using routing cost

The detailed algorithm is described as follows: In line 2, 3 every intermediate node forwards the request only to the neighbors that are closer to the source node $N_S$ than itself and farther away from the destination node $N_D$ to maintain the coverage and connectivity. In line 4, routing cost of each path from source to destination and intermediate nodes in updated. In line 5, the paths having the high cost is discarded according to the defined metric in equation (3). In line 6, each path is assigned the probability for successful transmission according to routing cost metric.

# 5. SIMULATION AND RESULTS

## 5.1 Simulation Environment

We have considered a stationary WSN of size 400×400 with a maximum transmission range of 60 m. The complete list of simulation parameters are described in Table 1. Other sources of energy consumption like sensing, processing, and idle listening are neglected. MAC-layer behaviors such as contention, duty cycles, or packet buffering are not addressed. We have simulated the proposed scheme on ns-2[29].

Table 1. Simulation Parameters

| **Parameters** | **Value** |
|---|---|
| Region radius under consideration | 400 m*400 m |
| Nodes sensing range | 60 m |
| Number of Nodes | 500 |
| Initial energy per node | 5 J |
| Network bandwidth | 2 Mbps/s |
| Power to run the transmitter/receiver circuitry | 70 nJ/bit |
| Power for the transmit amplifier to achieve an acceptable SNR (Signal to Noise Ratio) | 120 pJ/bit/m$^2$ |
| Size of a data packet | 4096 bits |
| Size of a control packet | 20 bits |
| Data transmission rate | 4096 bits |

## 5.2 Results
## Energy Consumption

Figure 8 presents the energy consumption of the proposed scheme with well known PEACH clustering protocol [22] and LEACH [14] when the maximum transmission range is 60 m. The results demonstrate that the energy consumption of proposed neural network based clustering is smaller than PEACH and LEACH. This is due to the fact that in the proposed scheme the routes are chosen according to the routing cost metric defined. This routing cost metric chooses only those routes those having least value of this metric and preserve the coverage and connectivity. Hence the mean residual energy in the proposed scheme is more than PEACH and LEACH, i.e., the proposed scheme prolongs the network lifetime compared to PEACH and LEACH.

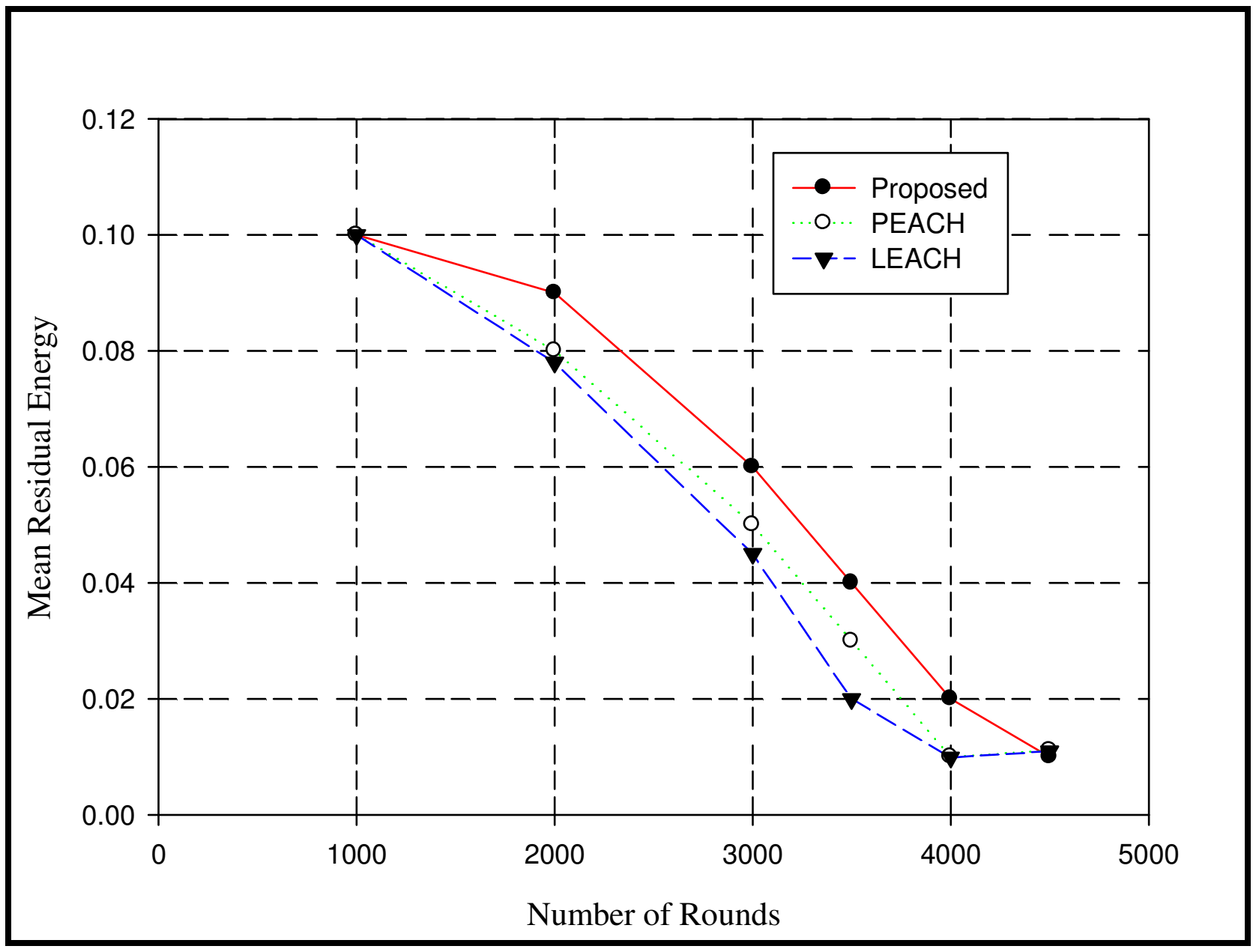

Figure8. Mean residual Energy in PEACH and Proposed Scheme

## Number of nodes with alive nodes percentage after finite number of rounds

Figure 9 presents the number of nodes alive when using clustering in proposed scheme, PEACH and LEACH. This result directly reflects the network lifetime of WSNs. In the case of networks using PEACH and LEACH with the maximum transmission range r = 60 m, where a node runs out of energy occurs nearly after 4000 rounds, while in propped scheme there is a slight improvement and node runs out of energy in nearly 4200 rounds. Again this is due to the routing metric considered in the proposed scheme. The percentage of alive nodes is presented in Figure 10. The percentage of alive nodes after a finite number of rounds is more in the proposed scheme than PEACH.

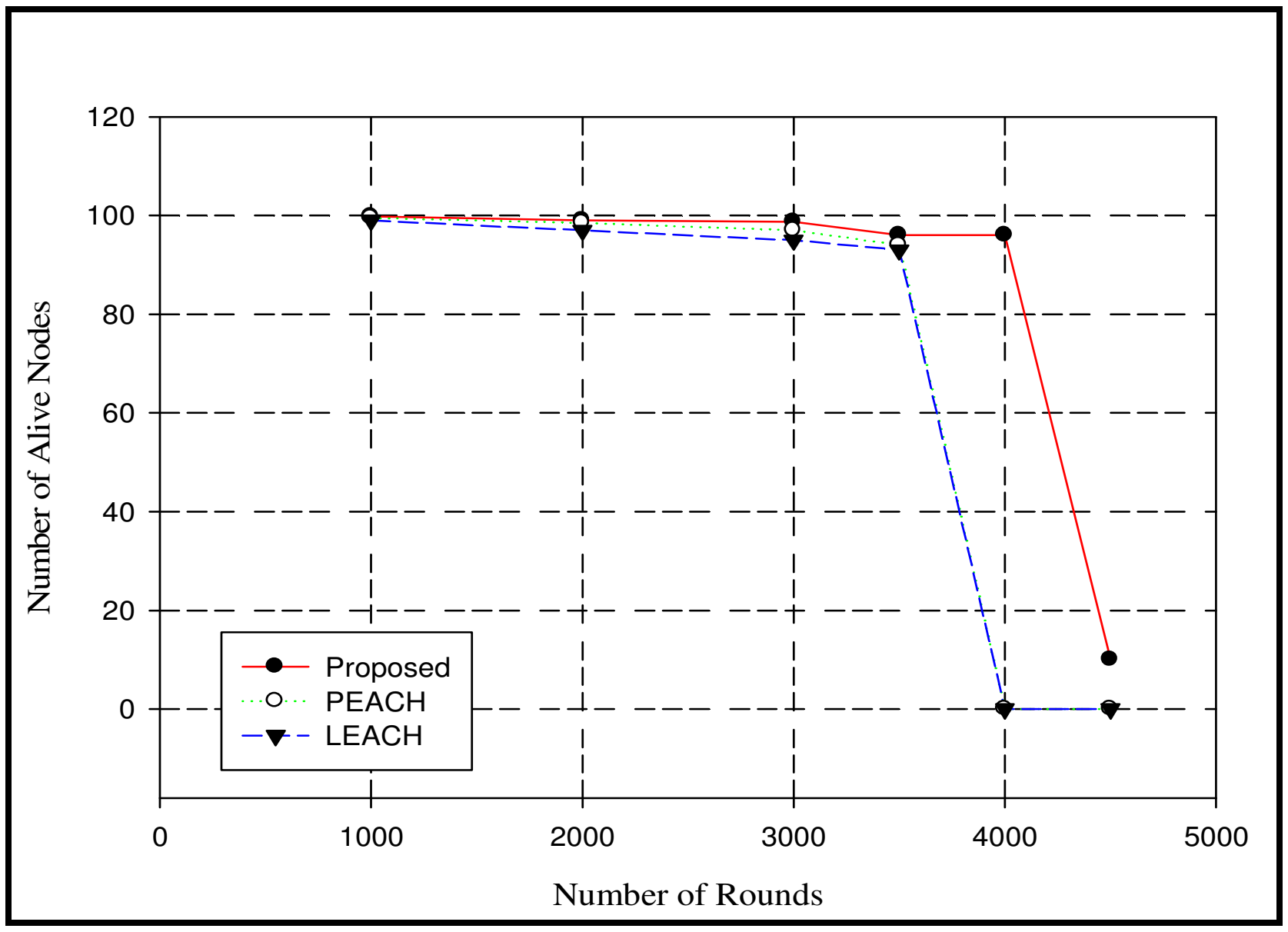

Figure 9. Number of alive nodes in PEACH and Proposed Scheme

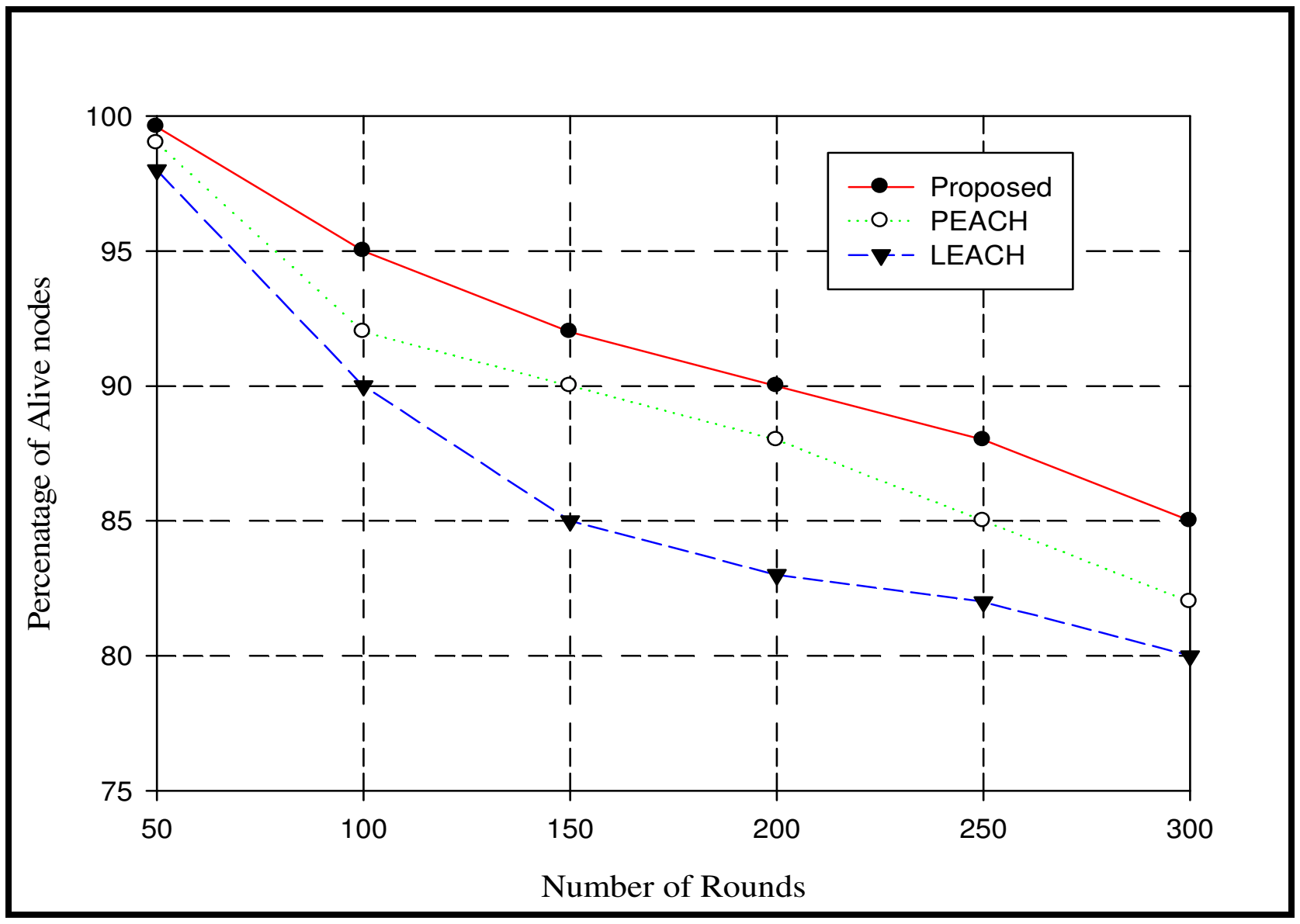

Figure 10: Percentage of alive nodes in PEACH and Proposed Scheme after 1500 rounds

## Impact on Network Coverage

We have also compared the proposed scheme with respect to network coverage ratio. It is measured as the coverage quality of the network, i.e. ratio of radio range and sensing range of the network. Zhang et al. have proved if the radio range is at least twice as large as the sensing range, the network coverage implies network connectivity [30]. That is, as long as the set of active nodes completely covers the monitored region, the network is connected. With a given the region radius R, the node sensing range and the node number N, we plot the network coverage ratio to get the average number of active nodes in the network in Figure 11. As shown in figure, with an increase in coverage ratio of the network, number of alive nodes decreases in all schemes. But in the proposed schemes numbers of alive nodes are more than other two schemes. This is due to the fact that nodes in the proposed scheme choose their routes according to the routing cost metric defined in equation 3. Hence a proper route is chosen and used during the whole process. Moreover the selection of CH is done by neuron having adaptive learning mechanism.

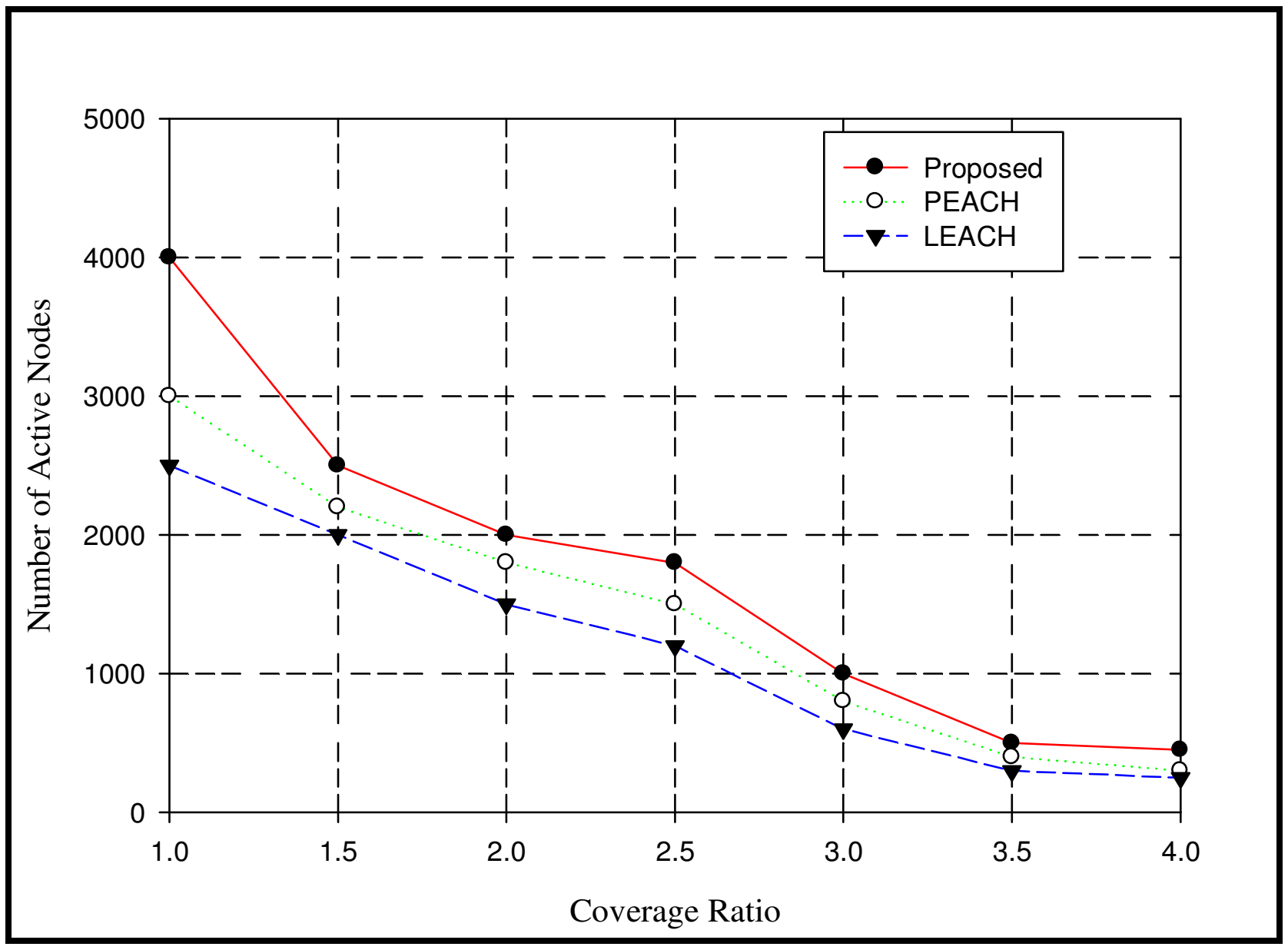

Figure 11. Number of active nodes with respect to Coverage Ratio

## Packet Delivery Fraction with network coverage ratio

The impact of network coverage ratio on packet delivery fraction is plotted in Figure 12. As shown in figure, the proposed scheme is able to have a packet delivery fraction of more than 95% with varying network coverage ratio compared to other schemes. This is due to the fact that routing paths are chosen depending upon the routing cost metric. As explained in the previous section, with an increase in coverage ratio, numbers of active nodes are more in the proposed scheme compared to other schemes. The numbers of active nodes have direct impact on packet delivery fraction. Hence packet delivery fraction also increases in the proposed scheme compared to other schemes.

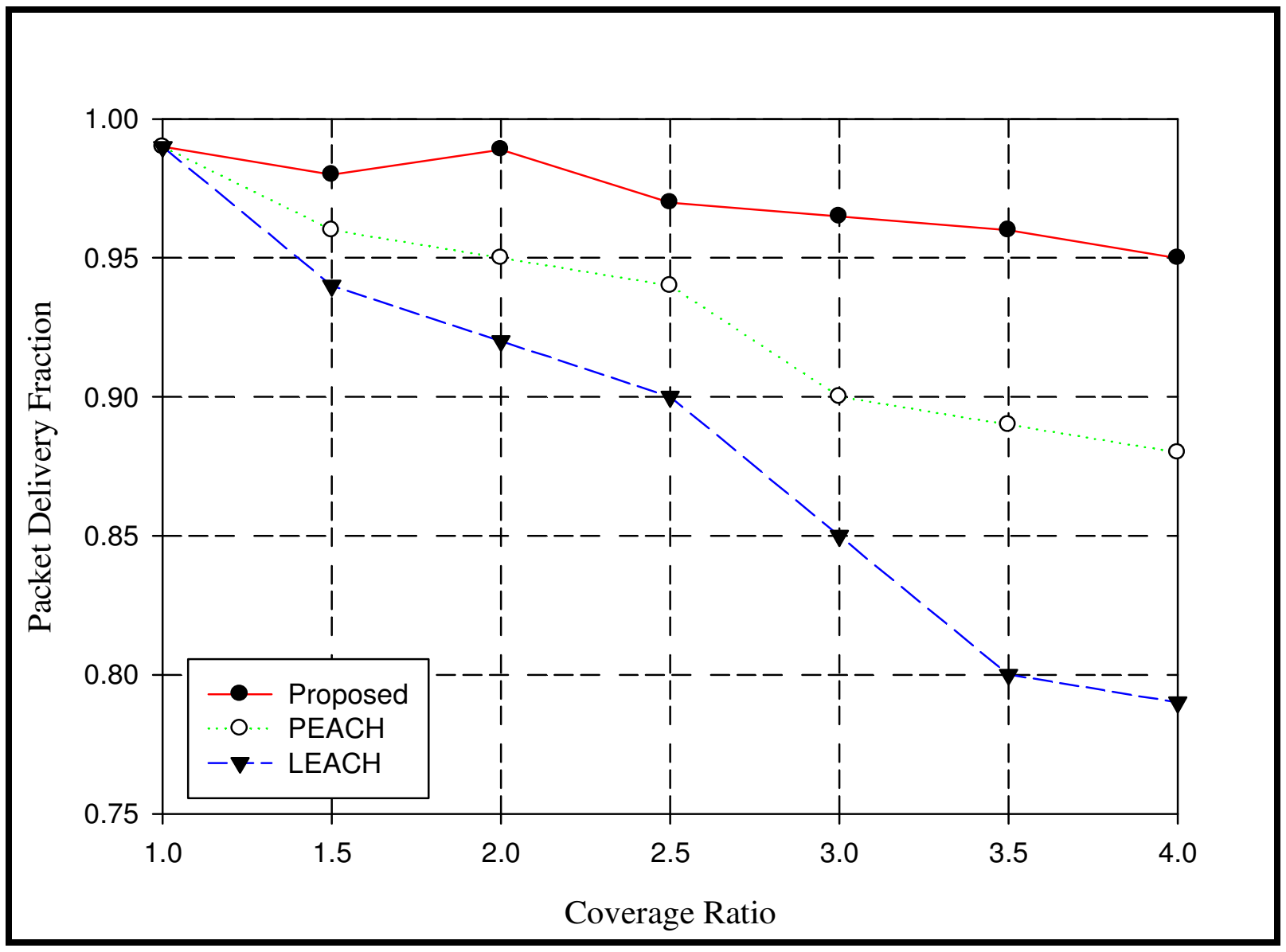

Figure 12. Packet Delivery Fraction with respect to Coverage Ratio

## 6. CONCLUSIONS

In this paper, we propose coverage and connectivity aware neural network based routing for WSNs. The problem is formulated as LP with specified constraints. The selection of CH is proposed using neural network with adaptive learning. The neurons are assigned weight according to the residual energy of the nodes in the network. A coverage aware routing metric is also included to choose the best route from the available ones. Once the routes are decided and one of the routes is chosen from these routes, the data transmission is performed using the defined metric. The performance of the proposed scheme is compared with PEACH and LEACH with respect to the metrics mean residual energy, percentage of alive nodes, packet delivery fraction and network coverage. The results obtained show that the proposed scheme is quite effective to deliver more than 95% of the packets to their destination with an increase in network coverage. Although with an increase in network coverage, the number of alive nodes decrease, but this decrease is less compared to other schemes. Hence the proposed scheme can be used in various scenarios with respect to coverage and connectivity.